\documentstyle[12pt,tables]{article}

\begin{document}
\begin{center}
{ \Large
Comment on ``Long-Ranged Orientational Order in Dipolar Fluids''\\}
\vspace{ 5 mm}
        M. Widom and H. Zhang\\
        Department of Physics, Carnegie Mellon University\\
        Pittsburgh, Pa. 15213\\
\end{center}

\begin{center}
\end{center}

	Recently Groh and Dietrich\cite{GD} calculated orientational
order transitions for dipolar fluids contained in spheroidal shapes of
aspect ratio $k$.  According to their study the thermodynamic state of
a dipolar fluid depends on the shape of the fluid's container. For
example, a homogeneous fluid in a short fat container would phase
separate when transferred to a tall skinny container of identical
volume and temperature. Their calculation thus lacks a thermodynamic
limit.

	Existence of a thermodynamic limit for dipolar fluids has
never been proven. Indeed, the long range anisotropic dipole
interaction presents significant difficulties. These include
mathematical difficulties such as the conditionally convergent, shape
dependent integrals examined by Groh and Dietrich.  Physically, the
long interaction range causes domain or texture formation to avoid
demagnetizing fields.  Still, Griffiths\cite{Gri} has proven the
existence of a shape independent thermodynamic limit for the free
energy of dipoles on lattices. So we presume that dipolar fluids also
possess a thermodynamic limit.

	The shape dependence in Groh and Dietrich's calculation
results from their assumption of spatially uniform magnetization. Real
crystalline ferromagnets form domains to avoid stray ``demagnetizing''
fields which arise when the magnetization is not parallel to the
sample boundary. In a liquid ferromagnet, however, the domain wall
thickness is limited only by sample size\cite{PinDeG}.  Thus the
magnetization is expected to rotate across the sample in a manner as
yet undetermined.  There is one sample shape which avoids the
demagnetizing field - the limit $k \rightarrow \infty$ corresponding
to an infinitely prolate spheroid. The bulk free energy calculated for
a uniform magnetization in this needle-like shape will match the true
free energy of any shape with appropriate spatially varying
magnetization.

	While the $k \rightarrow \infty$ limit is the uniquely
appropriate value for calculating bulk free energy density assuming a
spatially uniform magnetization, we do not expect a real ferromagnetic
liquid drop to take this needle-like shape in the absence of external
field.  The actual shape and magnetization texture of a ferromagnetic
liquid droplet remains an important unsolved problem.

	Taking $k \rightarrow \infty$ is not practical for computer
simulations. Recent computer simulations[4-7] solve the long range
interaction problem by combining Ewald summation with a reaction
field\cite{deL}. The limit of infinite surrounding dielectric constant
$\epsilon \rightarrow \infty$ cancels the demagnetizing field and
yields the true thermodynamic limit with a uniform state, while being
computationally more tractable than taking $k \rightarrow \infty$. In
simulations with $\epsilon = 1$, Weis and Levesque\cite{WL} find
domain formation consistent with our assertion that demagnetizing
fields create spatially nonuniform states.

	Now consider Groh and Dietrich's density functional theory
phase diagram in the $k \rightarrow \infty$ limit. This contains a
continuous magnetic transition at high temperatures and a tricritical
point below which phase separation occurs between dilute isotropic gas
and dense magnetized liquid.  Simulations of hard\cite{WL} and
soft\cite{WP,GS} dipolar spheres place the magnetized liquid state at
higher densities and lower temperatures than suggested by the present
calculation. A simulation of the Stockmayer fluid\cite{vLS} reveals
isotropic phase separation with a conventional critical point.  We
believe the density functional theory may be brought into closer
agreement with computer simulation by incorporating effects of random
particle positions which shift the magnetic transition to larger
densities\cite{ZW2}, and particle chaining\cite{DeGPin} which competes
with ordinary phase separation at strong dipolar coupling. Both
effects require three-body and higher terms in the density functional
theory.

	The phase diagrams found by Groh and Dietrich illustrate part
of a generic sequence of phase diagrams for dipolar fluids\cite{ZW}.
Many parameters, including the strength of the $1/r^6$ attraction in
the Stockmayer fluid, carry the phase diagrams through such a
sequence. But sample shape is {\it not} among such parameters.

	This work was supported by NSF grant DMR-9221596 and by the
A.P. Sloan foundation. We are indebted to R.B. Griffiths for numerous
discussions.


\begin{thebibliography}{99}

\bibitem{GD}
B. Groh and S. Dietrich, Phys. Rev. Lett. {\bf 72}, 2422 (1994)

\bibitem{Gri}
R.B. Griffiths, Phys. Rev. {\bf 176}, 655 (1968); M. Widom and R.B. Griffiths
(unpublished) recently extended this proof to cover a broad family of dipolar
fluids.

\bibitem{PinDeG}
P. Pincus and P.G. DeGennes, Solid State Comm. {\bf 7}, 339 (1969)

\bibitem{WL}
J.J. Weis and D. Levesque, Phys. Rev. Lett. {\bf 71}, 2729 (1993);
Phys. Rev. E {\bf 48}, 3728 (1993)

\bibitem{WP}
D. Wei and G.N. Patey, Phys. Rev. Lett. {\bf 68}, 2043 (1992)

\bibitem{GS}
M. Stevens and G. Grest, preprint (1994)

\bibitem{vLS}
M.E. van Leeuwen and B. Smit, Phys. Rev. Lett. {\bf 71}, 3991 (1993)

\bibitem{deL}
S. de Leeuw, J. Perram and E. Smith, Ann. Rev. Phys. Chem. {\bf 37}, 245 (1986)

\bibitem{ZW2}
H. Zhang and M. Widom, J. Magn. Magn. Mater. {\bf 122}, 119 (1993)

\bibitem{DeGPin}
P.G. de Gennes and P. Pincus, J. Phys. Kondens. Mat. {\bf 11}, 189 (1970)

\bibitem{ZW}
H. Zhang and M. Widom, Phys. Rev. E {\bf 49}, 3591 (1994)

\end{thebibliography}
\end{document}